# Influence of chemistry and structure on interfacial segregation in NbMoTaW with high-throughput atomistic simulations

Ian Geiger [1], Jian Luo [2], Enrique J. Lavernia [3], Penghui Cao [4], Diran Apelian [3], Timothy J. Rupert [1,3,4,*]

[1] Material and Manufacturing Technology, University of California, Irvine, CA, USA

[2] Department of NanoEngineering, University of California, San Diego, La Jolla, CA 92093, USA

[3] Department of Materials Science and Engineering, University of California, Irvine, CA, USA

[4] Department of Mechanical and Aerospace Engineering, University of California, Irvine, CA, USA

* trupert@uci.edu

**ABSTRACT**

Refractory multi-principal element alloys exhibiting promising mechanical properties such as excellent strength retention at elevated temperatures have been attracting increasing attention. Although their inherent chemical complexity is considered a defining feature, a challenge arises in predicting local chemical ordering, particularly in grain boundary regions with enhanced structural disorder. In this study, we use atomistic simulations of a large group of bicrystal models to sample a wide variety of interfacial sites (grain boundary) in NbMoTaW and explore emergent trends in interfacial segregation and the underlying structural and chemical driving factors. Sampling hundreds of bicrystals along the [001] symmetric tilt axis and analyzing more than one hundred and thirty thousand grain boundary sites with a variety of local atomic environments, we uncover segregation trends in NbMoTaW. While Nb is the dominant segregant, more notable are the segregation patterns that deviate from expected behavior and mark situations where local structural and chemical driving forces lead to interesting segregation events. For example, incomplete depletion of Ta in low-angle boundaries results from chemical pinning due to favorable local compositional environments associated with chemical short-range ordering. Finally, machine learning models capturing and comparing the structural and chemical features of interfacial sites are developed to weigh their relative importance and contributions to segregation tendency, revealing a significant increase in predictive capability when including local chemical information. Overall, this work, highlighting the complex interplay between local grain boundary structure and chemical short-range ordering, suggest tunable segregation and chemical ordering by tailoring grain boundary structure in multi-principal element alloys.

**Keywords:** Atomistic simulations, Grain boundary segregation, Multi-principal element alloy, Refractory complex concentrated alloy, Interface structure, Chemical short-range order

## I. INTRODUCTION

The rapid emergence of multi-principal element alloys (MPEAs) since 2004 marks a significant shift from centuries of conventional alloy design.[1] Whereas traditional alloys are composed of a base metal with relatively small amounts of alloying elements, MPEAs are multi-component metal alloys in equiatomic or near equiatomic proportions.[2–4] This paradigm shift uncovers new opportunities in compositional space with nearly limitless possibilities for elemental combinations;[5] several MPEAs have already demonstrated promising properties including excellent high temperature strength,[6–8] improved fracture toughness at cryogenic temperatures,[9] and outstanding oxidation and wear resistance.[10,11] Early studies aimed to correlate such properties to the stabilization of a single-phase random solid solution by high configurational entropy,[2] yet more recent work suggests that a diversity of microstructural landscapes is more the rule than exception, with different combinations of phases and microstructures promoted by altering the composition, temperature, and/or processing route. For example, iterative changes in the elemental concentrations of non-equiatomic FeMnCoCr MPEAs led to significant improvements in strength and ductility by altering deformation mechanisms from single-phase, dislocation-mediated plasticity to dual-phase, transformation-induced plasticity.[12–14] With potentially new or amplified mechanisms driving these enhanced properties, improved understanding and prediction of microstructure is necessary to enable continued advancements of MPEAs.

Grain boundaries are an ubiquitous microstructural feature that have been extensively studied in traditional alloys to better understand how structure and composition at interfaces affect both microstructural evolution and subsequent material properties.[15,16] In multi-component systems, the addition of even small amounts of dopants can significantly alter properties through grain boundary segregation, as shown in Al-Co,[17] Pt-Au,[18] and Mg-Gd[19] binary alloys. In Bi-doped Cu or Ni, for example, the formation of Bi bilayers at the grain boundaries drastically lowers the cohesive strength of the interfaces and transforms the generally ductile Cu or Ni into a brittle material.[20–23] Interfacial segregation is similarly observed in MPEAs, but is even more complicated as the inherent chemical complexity in elemental bonding often translates to complex grain boundary compositions. This behavior has been probed both experimentally[24–26] and predicted computationally,[27,28] again showing significant influence on mechanical properties. For example, using atom probe tomography, Ming et al.[26] showed that grain boundary decohesion via Ni, Cr, and Mn nanoclustering drives the loss of ductility in a CrMnFeCoNi MPEA. The substantial influence of interfacial state on material behavior underscores the need to better understand interfacial segregation in MPEAs as a critical step in improving material design.

The development of theoretical frameworks has been a principal focus of grain boundary engineering over the last century and has successfully uncovered the thermodynamic driving forces for segregation in simpler alloys.[29–31] Despite some progress, open questions remain about how such driving forces are expressed in chemically complex alloys. In many dilute alloys, atomic size mismatch can contribute to low bulk solubility, inducing segregation and simultaneously relaxing bulk and boundary stresses.[21,32] However, recent work from He et al.[33] showed that larger Bi and Pb atoms can segregate to compressed grain boundary sites at Mg coherent twin boundaries, thus indicating that chemical bonding can dominate over structural considerations even for low solute concentrations. Beyond dilute alloys and two-component systems, chemical interactions likely exhibit greater influence on segregation tendencies, requiring a thorough analysis. For MPEAs, the foremost chemical interactions that should be considered are often expressed in the bulk as elemental clustering (enthalpic favorability between two like elements) and chemical short-range ordering (CSRO) (enthalpic favorability between two unlike elements). These ordering tendencies have been observed to be prominent in several promising MPEAs.[34–36] Antillon et al.[36] performed Monte Carlo (MC)/molecular dynamics (MD) simulations of face-centered cubic CoFeNiTi to highlight different clustering and ordering tendencies at 700 K, 1100 K, and 1500 K, ultimately showing that chemical effects in the bulk are a strong function of temperature. Wynblatt and Chatain[37] simulated a CoNiCrFeMn alloy to decipher the role of chemical and structural driving forces at both surfaces and grain boundaries by adapting a two-component model that relates the enthalpy of segregation to: (1) interfacial energy of a solute atom, (2) elastic strain minimization, and (3) chemical interactions. These authors showed that excess Cr at the interface is the result of a strong affinity of Cr for itself, despite Mn having lower interfacial energy and greater reduction in elastic strain in high volume defect sites. Similar temperature-dependent ordering tendencies have been the focus of several studies on refractory, body-centered cubic (BCC) MPEAs that exhibit promising strength retention at elevated temperatures.[7,38,39] Furthermore, correlations between CSRO and dislocation propagation suggest that dislocation mobility, and therefore plastic deformation, is heavily influenced by an intrinsically rocky energy landscape.[40] Few studies to date have aimed to decipher how strong CSRO affects equilibrium grain boundary compositions, a critical component of microstructural stability in these refractory MPEAs.

In this work, high-throughput atomistic simulations are used to study local and global trends in segregation behavior of a BCC NbMoTaW refractory MPEA that exhibits a strong tendency for CSRO.[27,39,40] To probe a variety of structures and grain boundary sites, a dataset of 243 equilibrium and metastable symmetric tilt bicrystal models are generated and relaxed chemically via hybrid MC/molecular statics (MS) simulations. Our results show that Nb segregates most heavily, while Ta and W deplete almost completely, with Mo showing more subtle deviations away from the base equiatomic composition. To isolate segregation trends by elements, an effective coordination number is calculated that illuminates three distinct enrichment regimes within the overall spectrum of grain boundary sites. Of the three, the most compositionally complex regime corresponds to over-coordinated, compression sites. As effective coordination number increases, grain boundary sites tend to be filled primarily by Nb atoms, with Mo exhibiting some co-segregation behavior. Using a machine learning model that has been shown to accurately predict segregation in simple binary alloys, as well as a modified version with chemical descriptors, the relative importance of chemical and structural contributions on interfacial enrichment can be isolated, suggesting that strongly ordered atoms in the bulk can segregate to chemically favorable sites despite structural instability. Chemical and structural driving forces for each element are uncovered and discussed in the context of the CSRO observed in the bulk regions.

## II. Computational Methods

Atomistic simulations were performed with the Large-scale Atomic/Molecular Massively Parallel Simulator (LAMMPS) software package.[41] Visualization of atomic datasets was performed with the OVITO software,[42] with adaptive Common Neighbor Analysis[43] (aCNA) used for structural analysis and dislocation analysis performed using the Dislocation Extraction Algorithm (DXA).[44] A Moment Tensor Potential (MTP) developed for the NbMoTaW system was used to model atomic interactions.[40] This potential stems from a class of machine-learning interatomic potentials and accurately captures physical properties such as melting points for the four component elements close to those of experimental values, as well as important chemical properties such as segregation phenomena and CSRO, the latter of which has been similarly predicted in other studies.[27,45]

In this work, we quantify the bulk ordering tendencies of NbMoTaW at 0 K using a relaxed defect-free cube (Figure 1(c)) and the pairwise Warren-Cowley order parameter,[46] defined as:

$$\alpha_n^{ij} = 1 - \frac{p_n^{j,i}}{c_j}. \quad (1)$$

The order parameter, $\alpha_n^{ij}$, is an element interaction descriptor in which $i$ represents the center atom's element type and $j$ refers to an atom type present in the $n$th neighbor shell around $i$. $p_n^{j,i}$ is the concentration of atom type $j$ around type $i$ in the designated shell and $c_j$ is the global concentration of atom type $j$. The sign of $\alpha_n^{ij}$ indicates the favorability of an elemental interaction, with negative values representing increased probability of atom type $j$ in the $n^{th}$ shell, positive values representing a decreased probability in that shell, and values near zero indicating little correlation between $i$-$j$ pairs. For example, a tendency for like elements to cluster (i.e., $\alpha_n^{ii}$) is identified by negative values in both the first and second neighbor shells, as in the case of Nb-Nb in Figures 1(a) and (b). Similar first shell trends are observed for W-W pairings, however a positive value in the second shell suggests that W clustering over longer length scales is not favorable at 0 K. Ordering behavior between two unlike elements is determined by a negative value for $\alpha_1^{ij}$ and a positive value for $\alpha_2^{ij}$, and is observed most acutely for the Mo-Ta elemental pairing. This signal, paired with significant resistance to clustering in the first shell for their respective like-like interactions (i.e., large positive $\alpha_1^{ii}$), indicates a high probability of forming B2(Mo,Ta) CSRO and aligns with previous studies on this system.[27,39,45]

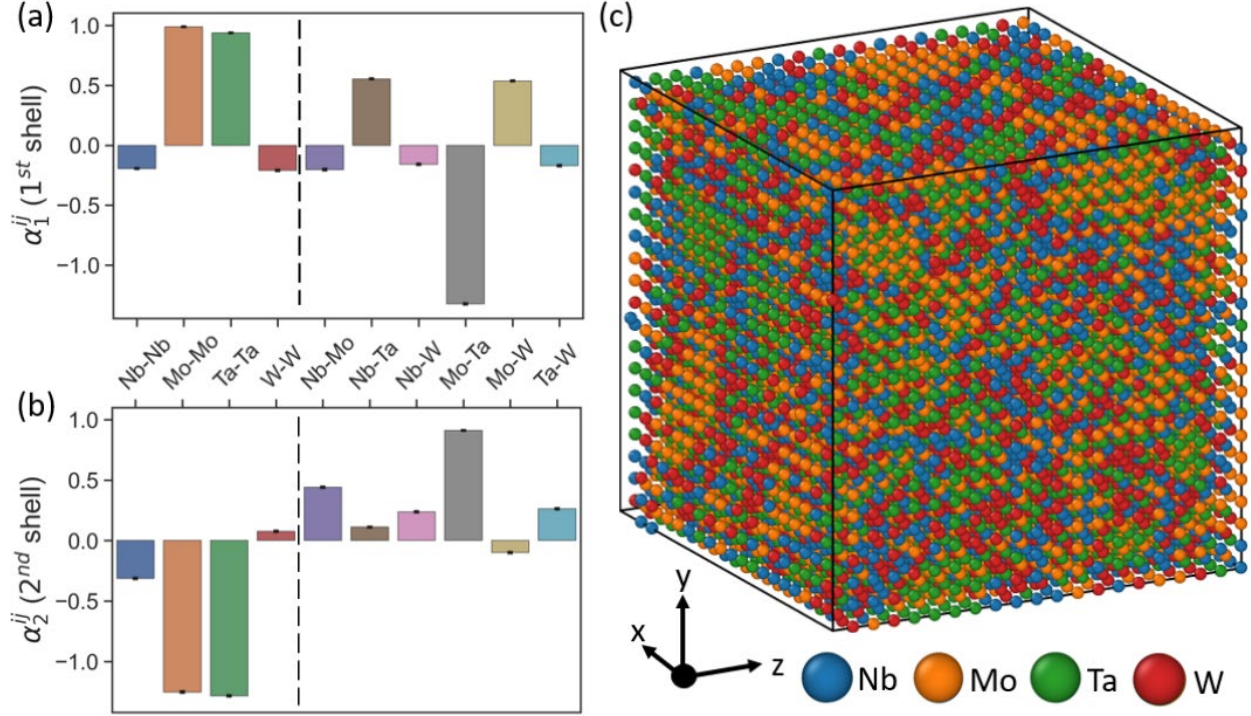

**FIG 1.** CSRO values, $\alpha_n^{ij}$, from the pairwise Warren-Cowley framework for (a) the first nearest-neighbor shell and (b) the second nearest-neighbor shell of single crystal NbMoTaW equilibrated at 100 K. Negative values indicate favorable interactions, while positive values indicate unfavorable interactions. The dashed line separates like-like (left) and unlike (right) elemental relationships. (c) The relaxed crystal from which the CSRO parameters were calculated, where distinct regions of Mo-Ta ordering of B2 type can be seen.

To sample a variety of grain boundary structures and unique atomic environments, bicrystal models were created from 81 tilt angles, ranging from 2.26° to 87.09°, along the [001] symmetric misorientation axis. Nb is used as the reference element when making the initial bicrystals because it has the largest atomic radius of the four elements in the quaternary system, as calculated from the single crystal MPEA sample, and is expected to segregate most heavily to the interfacial region.[27] The initial Nb samples were generated using an iterative method developed by Tschopp et al.[47] wherein a series of rigid body translations of two grains with fixed misorientation is performed, followed by successive atom deletion steps for increasing radial cut off values to avoid atomic overlap at the boundary and to generate interfaces with varying atomic densities. A conjugate gradient minimization is then performed to relax the final structure. Periodic boundary conditions were used in all directions resulting in two grain boundaries per cell with the minimum distance between interfaces selected to be 12 nm to eliminate the influence of one grain boundary on the other. In probing grain boundary structures when searching for a low-energy state, this grid search method passes through and outputs many higher-energy metastable configurations as well. While these high-energy configurations are often discarded if only the lowest energy configuration is targeted, metastable grain boundary states can also be of significant interest. Both simulations[48–50] and experiments[51] show that grain boundaries are not always in equilibrium due to high temperature conditions, dynamic processing, constraints from neighboring grains, and other considerations. In this work, we study both the lowest-energy state obtained at 0 K using this procedure, but also two metastable configurations per tilt angle to introduce a variety of additional grain boundary sites to investigate segregation behavior more thoroughly. Using green to represent BCC atoms and gray to represent grain boundary atoms, Figure 2(a) highlights an example of three configurations for a 62.9° tilt angle: (1) the lowest energy configuration, (2) the highest energy state, and (3) one intermediate energy structure.

Subsequent doping of each bicrystal was performed by randomly replacing Nb atoms with one of the other three elements until the overall composition of the cell consisted of 25% of each atom type (Figure 2(b)), resulting in a random solid solution. For the 243 bicrystals, simulation cell sizes range from 13,200 atoms at the smallest to 49,870 atoms at the largest, resulting in 130,000+ grain boundary sites overall. MC/MS was then performed to allow for chemical segregation to relax the cell. MS with the simulation cell pressure held constant at zero was used so that structural comparisons between the chemically relaxed alloy and the pure Nb grain boundary structures are possible due to retention of the initial grain boundary configuration. Local stress states and atomic volumes at Nb interfaces provide reference or initial conditions for the resultant segregation in the quaternary alloy. Relaxation with MD simulations would have led to significant restructuring of the boundary, meaning no comparison with the starting/reference state could be made. Moreover, while complete relaxation of the boundary is useful for finding the lowest energy states, this would limit the types of interfacial environments that are probed. MS using only a conjugate gradient minimization preserves the metastable states and ensures that a wider net is cast. For each MC step, the number of attempted swaps performed in each iteration equaled one-sixth of the total atoms in the cell. If an atomic swap would lower the system energy, it is always accepted. If the system energy would instead be higher after the swap, it is accepted probabilistically according to an Arrhenius law.[52] The MC acceptance criterion was performed at 100 K to allow the system to sample energetic states that may not be accessible only through downhill energy probing and to reduce the chances of becoming trapped in a local energy minima. The convergence criteria for a fully relaxed system were chosen based on grain boundary concentration since segregation phenomena are the primary target in this study. Once the boundary concentrations change by less than 1 at.% for all elements over 120 MC steps, the simulation cell is determined to have reached a chemically-relaxed configuration (Figures 2(c) and (e)). Distinct chemical differences are observed between the random solid solution and chemically-relaxed interfaces in Figures 2(d) and (e), where blue represents Nb, orange represents Mo, green represents Ta and red represents W.

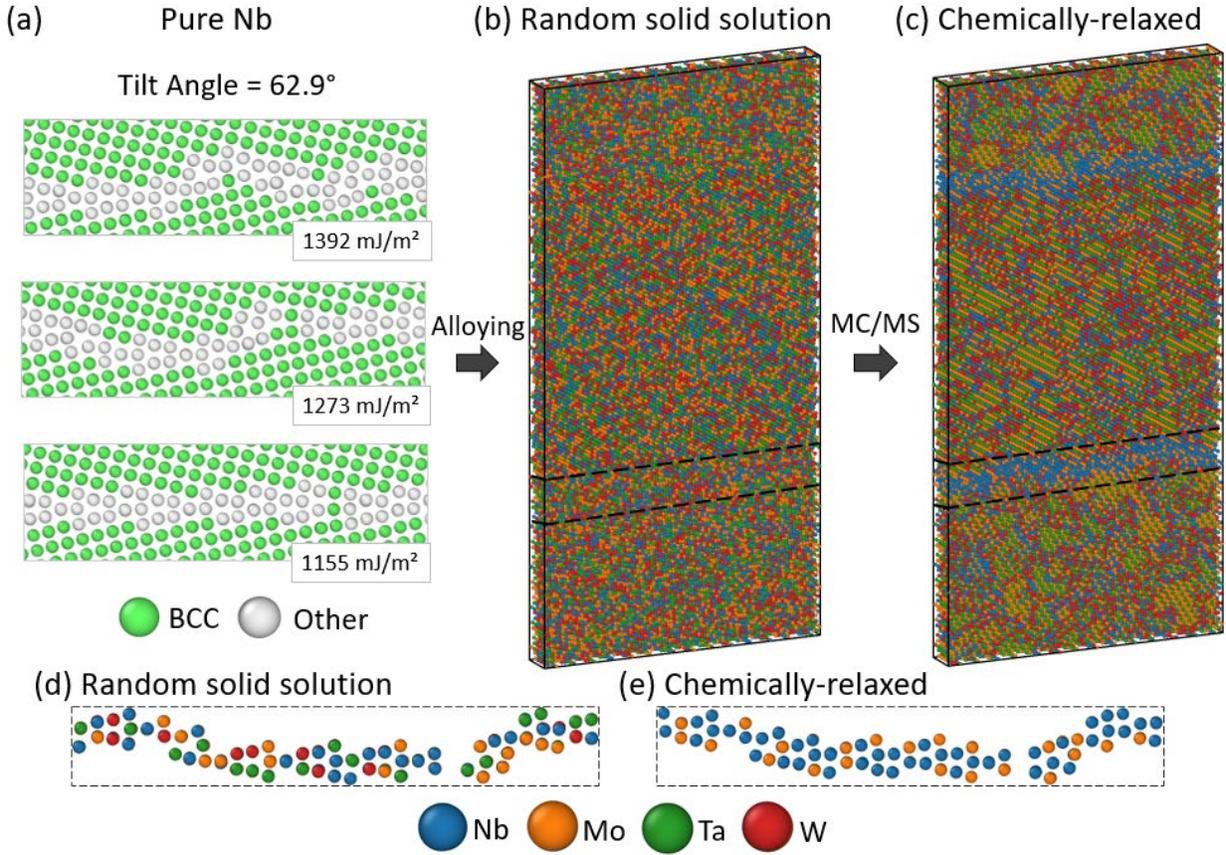

FIG 2. Example of boundary selection and subsequent doping for a tilt angle of 62.9°. (a) The selected pure Nb boundaries at different energies include the lowest energy (1155 mJ/m$^2$) and two metastable (1273, 1392 mJ/m$^2$) configurations. Atoms are color coded by structure type, where green represents BCC atoms and white represents grain boundary atoms. (b) Each bicrystal is doped to obtain equimolar concentrations of NbMoTaW in a random solid solution. (c) After chemical relaxation, the system is considered to be equilibrated. Clusters of blue at the interface indicate heavy Nb segregation. The interfacial (dashed) regions in (b) and (c) are highlighted in (d) and (e) to show compositional differences.

## III. RESULTS AND DISCUSSION

### A. Grain boundary sites and general segregation trends

Local hydrostatic stress is a physical parameter often used to predict segregation behavior in conventional alloys.[53,54] Since the MC/MS procedure constrains the grain boundary structure to the Nb reference state, we first identify key global and local structural features by analyzing the starting grain boundary configurations and then apply this understanding to the general segregation behavior observed in the MPEA. The complete set of grain boundaries in pure Nb is shown in Figure 3(a), presented as a function of tilt angle and with each point representing a different grain boundary configuration. Grain boundaries are often referenced as low-angle or high-angle in the context of the Read-Shockley model,[55] with 15-20° rotation from perfect lattice match between the two grains acting as the boundary between these two regimes. Low-angle boundaries are typically characterized by discrete dislocation arrays. In contrast, high-angle grain boundary structures exhibit more structural variation and are generally more disordered, except in "special" cases at specific misorientations.[56] In this work, we identify two critical transitions where important shifts in grain boundary structure occur. The first is a region denoted by the dashed line in Figure 3(a) at 35°, which extends slightly beyond the classical definition of low-angle boundaries. This first transition indicates where the lowest energy grain boundary structures transition from fragmented defects to a continuous planar defect, as shown by two insets on either side of this critical angle in Figure 3(a). A similar transition occurs at 61°, marking the beginning of a second region with fragmented grain boundary defect structures. Important connections to segregation behavior will be discussed in the context of these transitions in later sections.

Figure 3(b-d) highlights several grain boundary configurations in each angular region from the pure Nb samples. Adjacent to the grain boundary configuration colored according to structure type (obtained from aCNA) is the associated hydrostatic stress state and Voronoi volume for each atom. At low tilt angles, in the first disconnected boundary regime, the primary structural unit observed for the lowest energy configurations is the kite-shape, with BCC atoms filling the space in between kites (top row of Figure 3(b)). Different structural units are observed for the higher

energy, metastable boundary configurations (lower row of Figure 3(b)), but these units are also separated by BCC regions. The importance of this structure and the intra-plane BCC region on segregation will become apparent in the following sections. For the high angle disconnected boundary regime, discrete structural units are also observed and are separated by BCC regions, some of which again exhibit the kite-shaped structural units (lower row of Figure 3(c)). Corresponding atomic stress plots for the disconnected grain boundaries, particularly the configurations with kite structures, show significant stress fields extending into the nearby BCC regions, with periodic compressive and tensile regions due to a combination of <100> and ½<111> dislocations, as identified using DXA. The most compressed sites are at the kite tip, while the most expanded sites are at the back of the kite shape. Per-atom volumes show that highly compressed sites align with smaller atomic volumes, while those with high tensile stresses match those with higher volumes. Most importantly, as the distance between discrete defect structures is reduced, the stress fields begin to relax due to the overlap of tensile and compressive regions, resulting in atomic volumes closer to the bulk value. For the continuous boundaries between 35° and 61° tilt angles, the defective regions are uninterrupted and therefore cause smaller elastic stresses in nearby BCC regions, as shown in Figure 3(d). Not only are the stresses in the adjacent BCC planes normal to the interface lower, but the local hydrostatic stress for atoms in the boundary exhibit lower stresses, most notably less compression compared to that of the kite tips in the disconnected regions. Also notable is that the continuous boundaries have many atomic sites with large Voronoi volume.

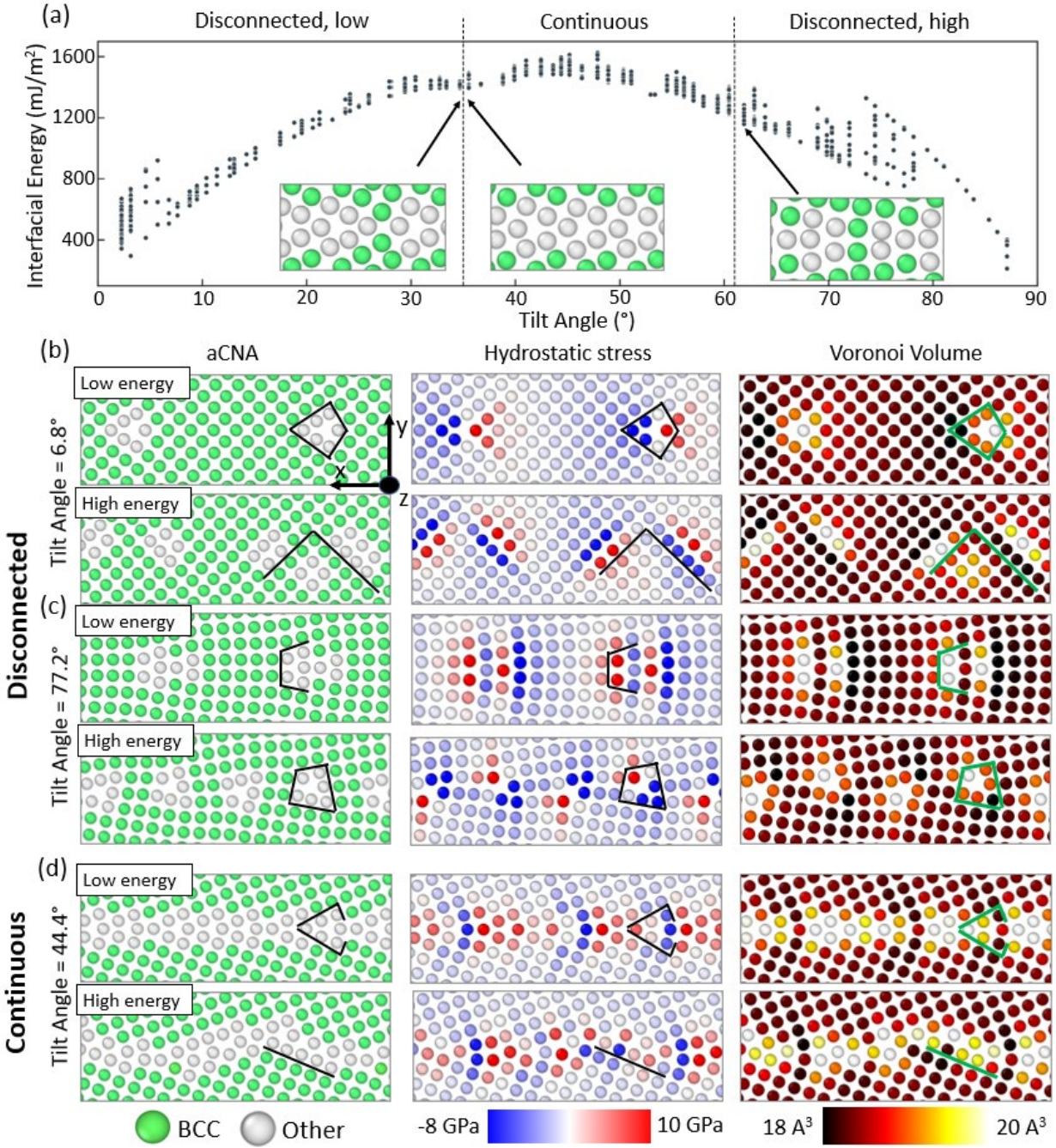

FIG 3. (a) Interfacial energy plotted as a function of tilt angle, showing low-energy and metastable grain boundary configurations for a given tilt angle and with dashed lines highlighting the transitions between disconnected and continuous boundary structures. (b), (c) Grain boundary structure, local hydrostatic stress, and Voronoi volume for representative boundaries from the two disconnected defect structure regimes. (d) Grain boundary structure, local hydrostatic stress, and Voronoi volume for a representative boundary from **the continuous defect structure regime. Black and green lines are used on each subplot to easily visualize similar grain boundary units across the three perspectives. Green and gray atoms refer to BCC and defect atoms, respectively.**

We next look to broadly convey how these interfacial features impact grain boundary composition in the relaxed MPEA. However, while the hydrostatic stress and Voronoi volume are highly correlated in the Nb grain boundary structures, local atomic distortions in the MPEA make the relationship between these two parameters less obvious. Therefore, we draw connections between local structure and composition by referencing both the initial Nb and final MPEA structural states. Importantly, these comparisons provide clear evidence for how the grain boundary structure impacts segregation behavior in and near the defect sites. Figure 4 presents several perspectives of one disconnected and one continuous defect structure boundary in the MPEA and pure Nb samples at 11.4° (Figures 4(a-f)) and 47.9° (Figures 4(g-l)) tilt angles, respectively. Figures 4(a-c) show the local atomic structure as well as the hydrostatic stress and Voronoi volume of each atom for the pure Nb model. Similar to the Nb boundary at 6.8° shown in Figure 3(b), the stresses and atomic volumes along the grain boundary plane also exhibit noticeable variations, although they are reduced in length due to closer proximity of the structural units. In Figures 4(d) and (f), the per-atom hydrostatic stress and Voronoi volume are presented for the same boundary in the MPEA. It is important to note that though the same parameters are used to compare the same boundary in the two metals, the magnitudes of the stresses and volumes in the MPEA are generally larger than in Nb. While volumetric trends are largely unchanged from pure Nb to the alloy, only the most stressed sites at the front and back of the kite structure are easily discerned in the MPEA (Figure 4(d)). The primary causes for this are local relaxations due to segregation and severe lattice distortion in the bulk regions, which make local trends in the boundary plane less dramatic. Despite this, patterns in elemental groupings at the interface are identified that demonstrate how the differences in boundary structure have a clear effect on local chemistry in Figure 4(f), with more elemental variation in the encircled, compressed regions. The gray arrows highlight the kite tip sites and a different element is observed for each kite unit. When either Mo (orange) or Ta (green) are in this site, local chemical ordering in the nearby BCC region (denoted by black circles) is observed, as indicated by alternating Mo and Ta atoms. In contrast, there is no chemical ordering expressed adjacent to the kite tip occupied by W despite similar magnitude of compression and atomic volumes. In the expanded regions on the right side of the kite and even in the nearby BCC regions on that side, Nb is the dominant segregating species. Figure 4(g-l) presents a representative continuous defect structure boundary. Similar relaxation in the boundary plane is observed from the pure Nb to the MPEA samples (Figures 4(h) and (j)), while trends in atomic volume are retained (Figures 4(i) and (k)). Although the interfacial structure becomes more complex, the range of local atomic volumes for the grain boundary sites is actually reduced, with notably fewer low volumes sites as compared to the disconnected boundary in Figure 4(a). In addition, the compositional complexity is reduced in the grain boundary (Figure 4(l)), where Nb acts as the primary segregating species and Mo acts as a secondary segregant.

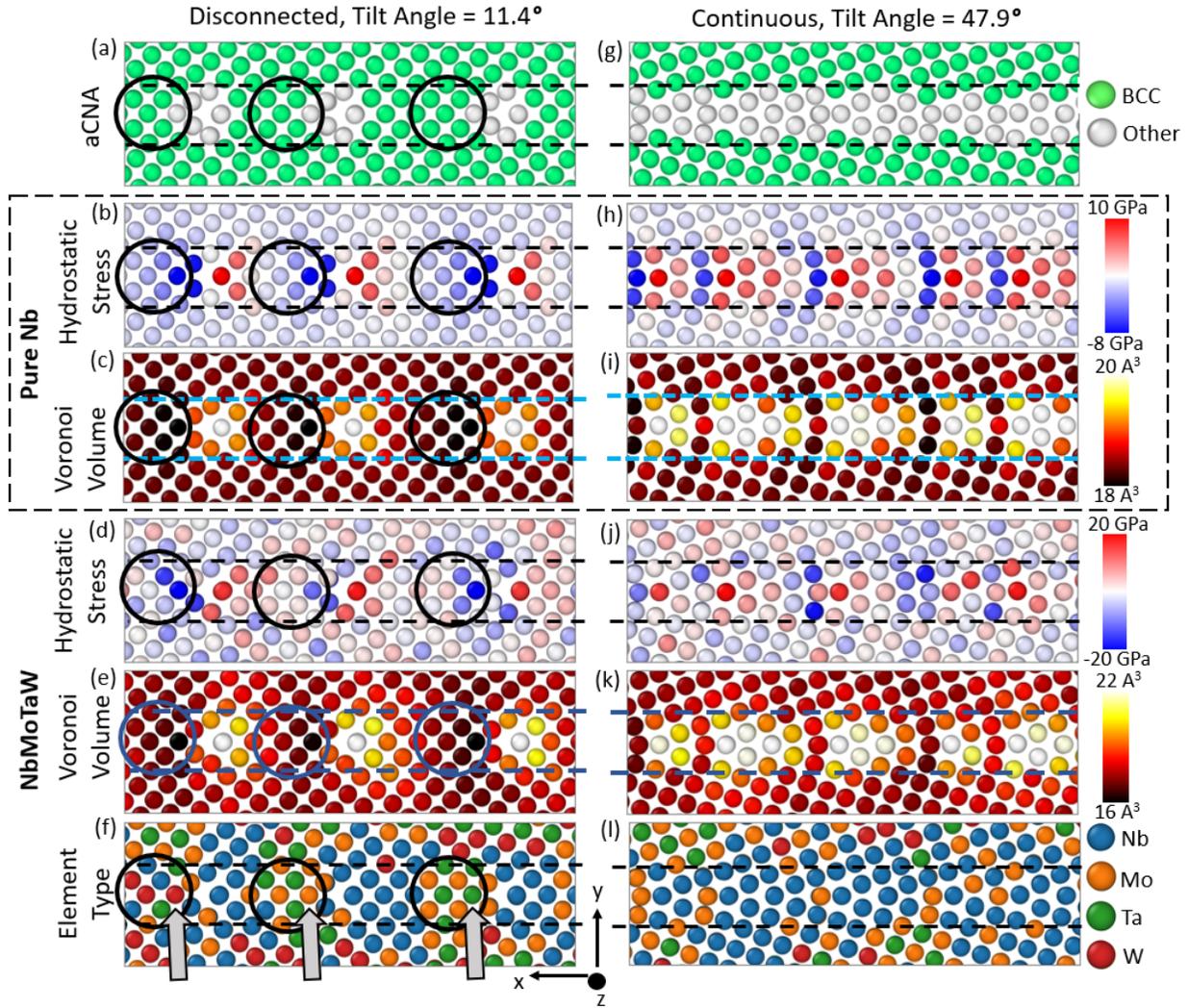

**FIG 4.** Two grain boundary configurations in the MPEA from the 11.4° and 47.9° tilt angles, colored according to (a), (g) aCNA, (b), (h) Voronoi volume from pure Nb, (c), (i) local hydrostatic stress from pure Nb, (d), (j) local Voronoi volume from MPEA, (e), (k) local hydrostatic stress from MPEA, and (f), (j) boundary composition from MPEA. The dashed lines indicate the grain boundary plane in each perspective. The circled regions in (a-f) highlight the BCC regions next to the kite tip (black arrow).

The local composition of two disconnected and two continuous boundaries is shown in Table I. For each boundary, the concentrations of each element are computed by either (1) including only defective grain boundary atoms (~0.6 nm in width) or (2) by including all atoms in a 1 nm thick slice at the grain boundary plane. The latter encompasses both the defect sites plus some BCC atoms nearby, primarily from the regions between disconnected defect units. In both disconnected boundaries sampled in Table 1, inclusion of the atoms from the 1 nm slice results in significantly different compositions than their counterparts that include only defect atoms, with the concentration of W doubling and Ta increasing even more so. The primary cause for this effect is the regions between defect units are not identified as grain boundary atoms yet are impacted by both the segregation to the defect regions and the local structural variation. For the continuous boundaries, the Ta and W concentrations remain very low, with only small increases due to the inclusion of a few near-boundary atoms perpendicular to the boundary plane. This analysis, along with that of Figure 4, shows that interfacial composition in this MPEA is intimately tied to grain boundary structure. The influence of that structure, especially for the disconnected boundaries, can extend beyond only the defect atoms by promoting clustering and ordering in the elastically strained regions in the nearby bulk. While it was useful to look at different ways of defining the grain boundary region here, we identify grain boundary atoms as only those which are defects for the remainder of the paper.

|  | **Disconnected** | | | | **Continuous** | | | |
|---|---|---|---|---|---|---|---|---|
| **Tilt Angle** | **6.8°** | | **10.4°** | | **47.9°** | | **60.5°** | |
|  | Defect atoms | 1 nm slice | Defect atoms | 1 nm slice | Defect atoms | 1 nm slice | Defect atoms | 1 nm slice |
| **at.% Nb** | 60.4 | 44.7 | 55.6 | 52.3 | 79.3 | 68.5 | 71.9 | 68.0 |
| **at.% Mo** | 28.3 | 25.0 | 32.5 | 22.1 | 20.1 | 26.4 | 27.6 | 26.9 |
| **at.% Ta** | 2.9 | 14.6 | 3.9 | 9.3 | 0.0 | 1.7 | 0.0 | 2.4 |
| **at.% W** | 8.3 | 15.8 | 8.1 | 16.3 | 0.6 | 3.3 | 0.5 | 2.7 |

**TABLE I.** Representative boundaries indicating compositional differences between disconnected and continuous type boundaries when including defect-atoms only and all atoms within a 1 nm slice about the grain boundary plane.

Compiled information on grain boundary composition is presented in Figure 5 as a function of tilt angle. At a particular angle, the three data points for each element represent the three boundaries investigated. While significant differences in elemental concentration for a given angle are observed in some cases (e.g., variations of approximately 10 at.% W in the low-angle, disconnected region), each elemental concentration profile follows a consistent trend. Nb enriches heavily for all tilt angles, with maximum values at both very low and intermediate angles. Mo enriches in the disconnected boundary regimes and depletes in the continuous boundary region, yet the Mo concentration at the interface generally remains within ±10% of the equiatomic bulk value. Ta and W deplete significantly at all angles, although incomplete depletion is observed for the disconnected boundaries.

Generally, these trends match observations made by Li et al.[27] for a polycrystalline model, although these authors reported more depletion of W than Ta. These differences could be the result of several factors, including differences in interatomic potential used, temperatures probed, and the sampled grain boundary types. However, our findings present clear trends not readily available from a polycrystalline grain boundary network as well. The changes we observe as a function of tilt angle suggest that opportunities for tailoring interfacial composition through grain boundary structure in this MPEA exist. The vertical dashed lines in Figure 5 again indicate the angles at which structural transitions between disconnected and continuous structure regimes occur. These transitions match the relative depletion of Mo as well as the complete depletion zone for both Ta and W. From this we can infer two important driving forces impacting segregation behavior, one or both of which dominate specific sites for disconnected boundaries. First, structural changes at these transition angles mark a loss of important grain boundary sites (i.e., compression sites) to which certain elements segregate favorability. Second, the loss of ordering/clustering within the grain boundary plane reduces the effect of the nearby BCC sites on the chemistry of the defective sites.

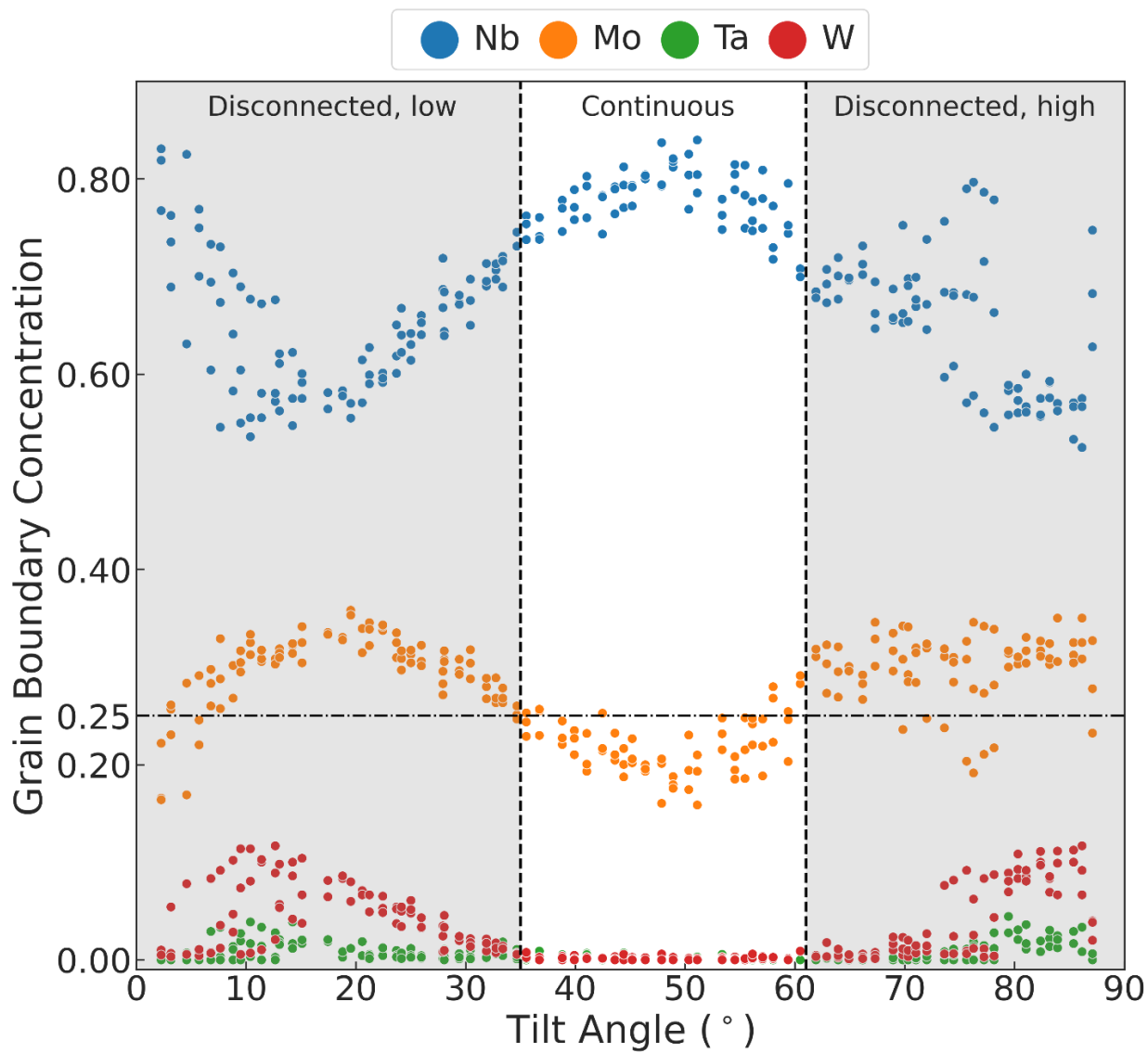

**FIG 5. Element concentration at the grain boundary sites as a function of title angle. For a given angle, each element has three data points corresponding to the lowest energy configuration and two metastable ones. The horizontal dashed line indicates the equiatomic fraction, while the two vertical dashed lines correspond to the boundaries between disconnected (gray shading) and continuous (white background) grain boundary structures.**

## B. Classification into grain boundary site types

Next, we aim to isolate the driving forces for grain boundary segregation on a per-element basis by first parsing all boundary sites into distinct local structural units and site types, and then analyze their local chemical profiles to explain this behavior. Since local stress and volume are both closely tied to element type with significant sensitivity to neighbor interactions, another descriptor is needed to sort grain boundary site types. Coordination number is commonly used to describe local bonding and geometrical environments in both grain boundaries and bulk environments,[33] and can be implemented in such a way that small local perturbations have little effect on the calculated value.[57] Calculations of the traditional (i.e., discrete) coordination number rely on relatively undistorted and symmetric environments, unlike those often found at an interface, as a result of a constant imposed search radius. For example, in the bulk, a radial cut-off of 3.02 Å and 4 Å, identified from the radial distribution function tool in OVITO, accurately calculates the BCC coordination number with 8 and 6 neighbors in the nearest and next-nearest neighbor shells, respectively. In the grain boundary, this method becomes unreliable due to local lattice distortion. Instead, we utilize here an effective coordination number, $N_c^*$, as implemented by Huang et al.[58] and Huber et al.[57] that calculates and compiles the neighbor coordination contributions as a function of distance from the central atom. The effective coordination number calculates a weighted coordination as:

$$N_c^* = \Sigma_j f(r_{ij}) \qquad (2)$$

where the summation term accounts for all atoms in the cut-off at a distance, $r_{ij}$, from the $i^{th}$ atom. The function in the summation is a Fermi-Dirac distribution calculated as:

$$f(r_{ij}) = \frac{1}{1 + \exp(\frac{r - r_0}{\sigma})} \qquad (3)$$

where $\sigma$ is the smearing parameter that ensures the coordination values have a continuous distribution and $r_0$ is half the distance between the first and second neighbor shells (3.025 Å). The chosen value for $\sigma$ (0.09 Å) is approximately 20% of the distance between the averaged first nearest neighbor and second nearest neighbor distances. Both values were selected following the procedure from Huber et al.,[57] and result in a calculated effective coordination in the bulk of approximately 8.4.

In Figure 6(a), the effective coordination number for each defect atom from the entire simulation dataset is plotted against the potential energy for that atom. Three distinct structural types can be identified, for which different elemental groupings become apparent, with the element counts for each type provided in Supplementary Note 1. Type 1 sites are characterized by considerable over-coordination and account for 5% of the total sites in the dataset. Type 3 sites are highly under-coordinated and account for 34% of the total sites. Type 2 site coordination values fall between these two extremes and are the most prevalent. Examples of the three structural sites are provided in representative examples of disconnected and continuous boundaries in Figure 6(b). The disconnected boundary provides the clearest visual distinction between site types, in which Type 2 sites are bounded by compressed sites (kite tip) corresponding to Type 1 and expanded sites (back of the kite) corresponding to Type 3. The continuous boundary is noticeably absent of Type 1 sites and is more abundant in Type 2 sites. In this boundary, the majority of Type 3 sites are found at or near the center of the grain boundary plane.

For Type 1, all elements are present with W being observed most frequently, accounting for 46.1% of the atoms. The segregation of W and Mo to compression sites can reduce high local elastic strains and be explained by purely structural driving forces, as these are the smaller elements of the four. The same logic cannot be applied to the larger Nb and Ta atoms, which together account for about 23% of atoms in these sites, suggesting that chemical interactions must play an important role. Type 2 sites are primarily composed of Nb and Mo atoms (~56% and ~43%, respectively) with inclusion of a few W and Ta atoms. In fact, the majority of all Mo and Nb grain boundary atoms are observed in this type of site. In under-coordinated, Type 3 sites, Nb is almost exclusively present. Under-coordination correlates with fewer bonds, meaning expanded sites and larger volumes. Purely elastic considerations would suggest that Nb and Ta would compete for these sites, however, only 5% of the total 756 Ta atoms are observed in Type 3. This is likely due to chemical interactions in the grain boundary that strongly favor Nb-Nb clustering.

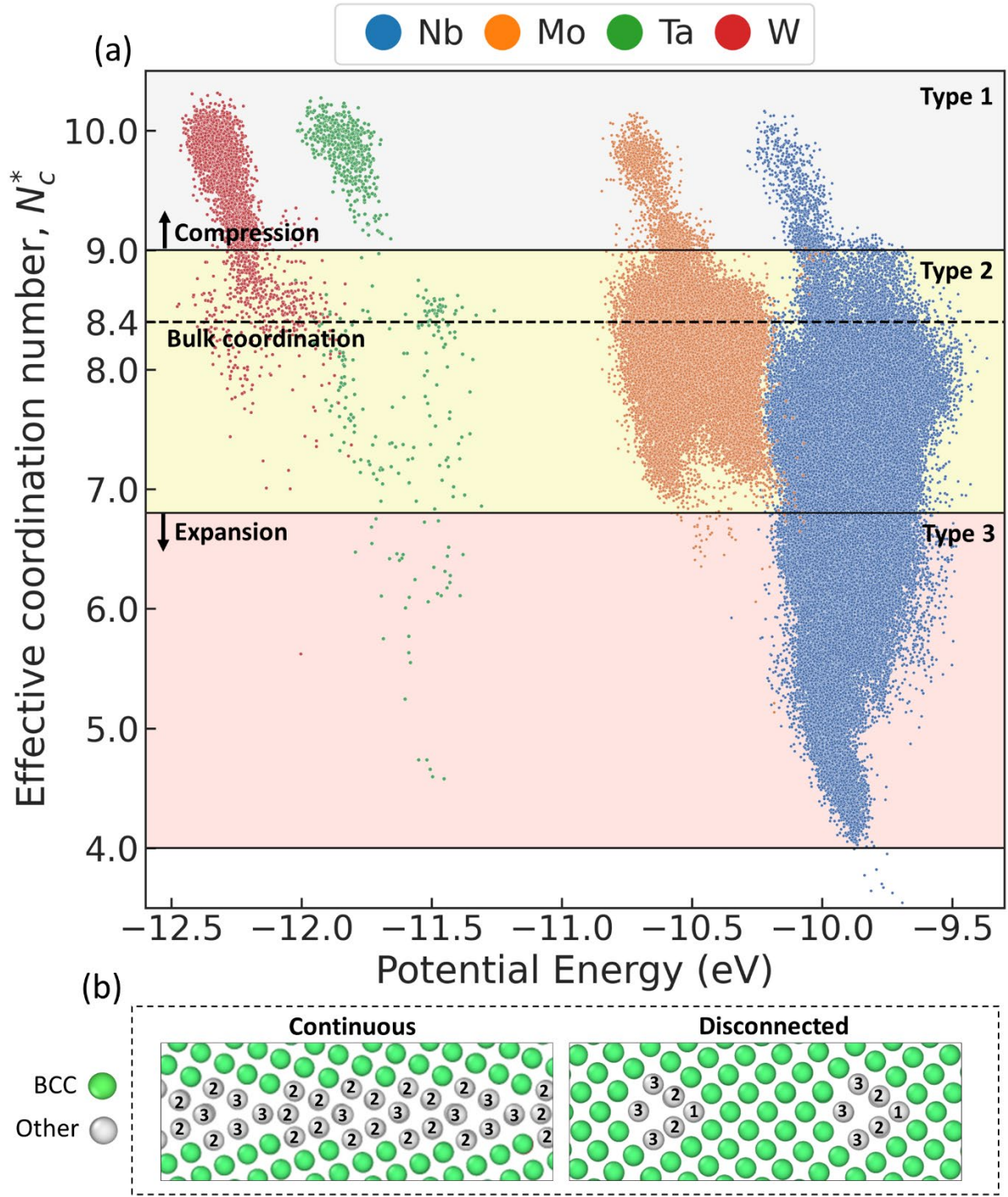

FIG 6. (a) Effective coordination number, $N_c^*$, versus potential energy for all grain boundary atoms in the dataset. The average $N_c^*$ in the bulk regions is 8.4 and is denoted by a horizontal dashed line. Three different grain boundary site types are identified, which vary by the relative compression (over-coordination) or expansion (under-coordination) compared to the bulk average. Data points are color coded by element type.

**(b) Two grain boundary structural units are shown to illustrate common local structural environments pertaining to each site type. Type 1 indicates compression site and type 3 denotes expansion.**

To better understand how these defect types align with previously introduced structural descriptors, each site type from the coordination analysis is further assessed by misorientation angle and local atomic volume. In Figure 7(a), the three types are considered as a whole (i.e., without separation by element type) as a function of tilt angle using a histogram plot normalized such that the summation of all bar heights equals 1. A kernel density estimation is provided as a line for each type to indicate angular trends in each distribution. While Types 2 and 3 are most prominent in the continuous boundary regime at intermediate tilt angles, site Type 1 is only observed in the two disconnected boundary regimes at very low and very high angles. The defect site types are further analyzed in terms of atomic volume in Figure 7(b) where only the kernel density estimation is shown, rather than the raw data, to enable visualization. In this figure, distributions for each of the four elements are shown separated by coordination type. For Type 1 sites, a pronounced bimodal distribution is observed for Mo and W. Nb shows a similar trend but with clear preference for relatively larger volumes, while Ta is only observed at relatively smaller volumes. The range of Type 1 atomic volumes is an indicator of differing magnitudes in stress relaxation of the compressed sites. Kite structures that are widely spaced have more compression at the kite tip than those with small separation, resulting in lower atomic volumes. As shown previously, wide spacing between grain boundary structural units can enable CSRO to evolve in BCC sites along the grain boundary plane, ultimately facilitating segregation via strong chemical interactions for otherwise heavily depleted elements like Ta. Histograms of element counts by atomic volume in Type 1 sites are provided in Supplementary Note 2 to provide more detail. In Figure 7(b), Type 2 Nb and Mo atoms are represented by a relatively symmetric single peak distribution in atomic volume, both centered at a value greater than the bulk average of 16.96 Å$^3$. Significant segregation of Mo to Type 2 sites may be considered unusual in that structural relaxation is likely not the primary driving force. While Nb is the largest atom in this alloy, Mo is one of the smallest, meaning that chemical interactions contribute significantly to the large quantities of Mo observed in these sites. An analysis of local chemical profile for each element and coordination type confirms this hypothesis in the following section.

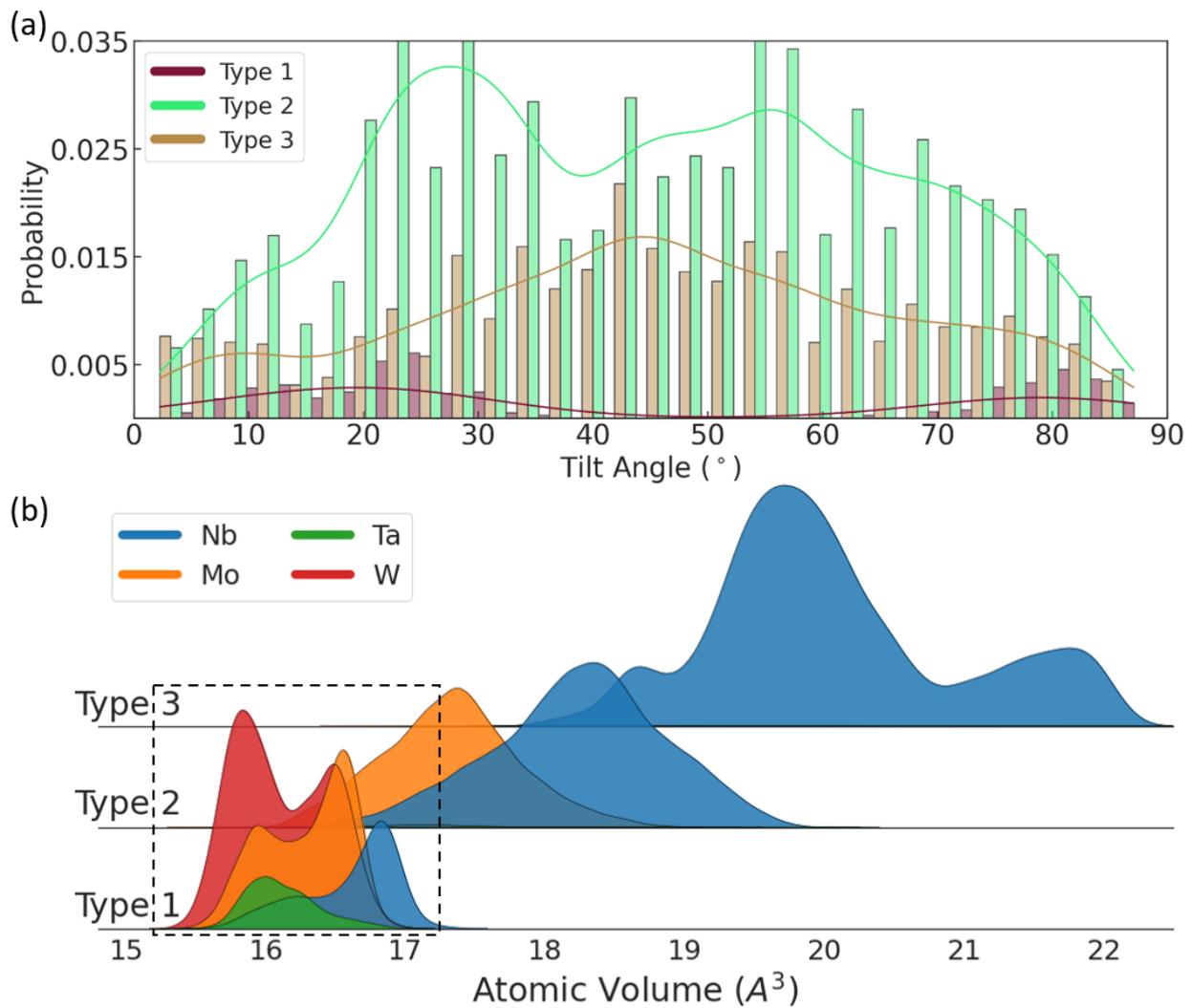

**FIG 7. (a) The totality of all atoms in each Region are plotted by misorientation angle. For each site Type, a kernel density estimation indicates the distribution trend. (b) Probability density functions illustrate trends in**

**atomic volume for each element in respective Types. The individual histograms of different atoms (Nb, Ta, Mo, and W) in the dashed box are explicitly shown in Figure S1. Note the Type 3 sites are predominantly occupied by Nb atoms.**

To elucidate local chemical environments encompassing a central/target atom, the concentration of neighboring elements within the effective coordination shell is first calculated. Here, we normalize the contribution to the effective coordination from a neighboring element atom type (e.g., the contribution from all neighboring Nb atoms) by the total effective coordination to get a local chemical profile for each defect atom in the dataset. An example is shown in Figure 8(a), where the central and neighboring atoms are labeled. For each effective coordination regime, concentrations of nearest neighbors for a central atom type are then averaged to highlight critical differences in local composition for different central atoms using heatmaps presented in Figure 8(b). On each figure, the $x$-axis contains the choice of central atoms and the $y$-axis shows the neighboring elements in the coordination shell.

Despite general structural similarity, Figure 8(b) highlights significant variance in local neighbor profiles within a coordination type that emphasize the influence of chemical preference on defect atoms. For example, the average concentration of Mo neighbors surrounding a central Type 1 Ta atom is 0.72, the highest value of all pairs in both Type 1 and 2 sites. In comparison, Mo neighbor concentrations around other Type 1 central atoms are as low as 0.25, suggesting that the Mo interaction is critical to favorable segregation for Ta. In Type 2 sites, a central Mo atom exhibits similarly high magnitude of partiality towards Nb neighbors. This is not observed for Type 2 central Nb atoms, further strengthening the hypothesis that Mo enrichment is reliant on Nb co-segregation.

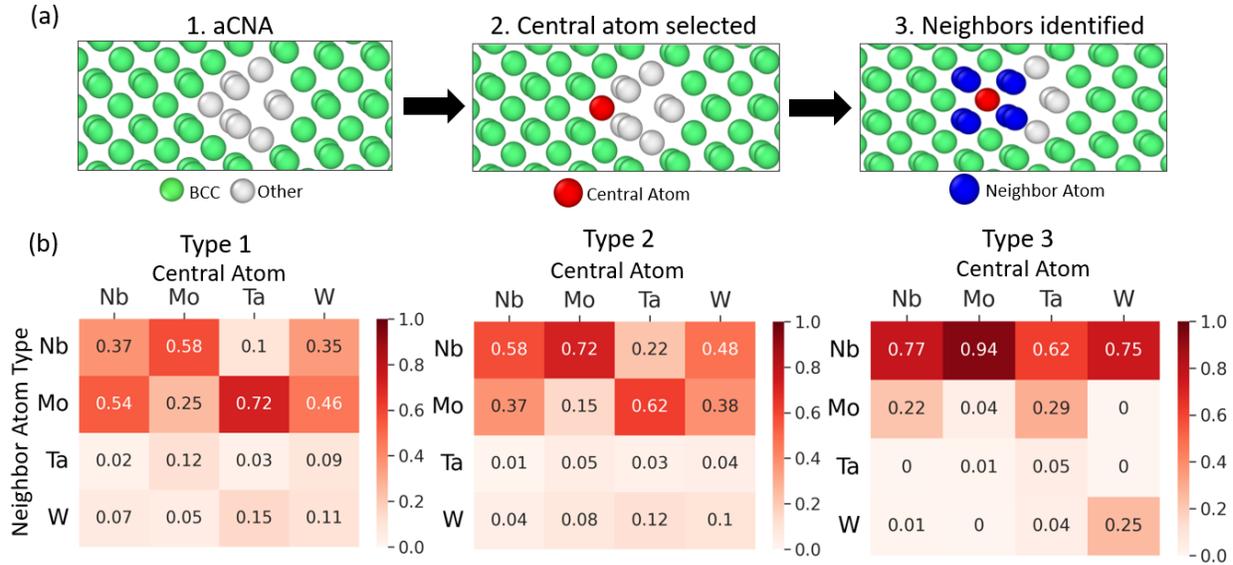

**FIG 8. (a) Visualization of the neighboring atoms selected around a central defect atom. (b) Heatmaps for each site type show the average concentrations of various elements in the nearest-neighbor shell (*y*-axis) around a central element type (*x*-axis).**

The local compositions around central atoms are shown in Figure 9 as a function of atomic volume, organized by site type. The line colors on each plot in Figure 9 represent the neighboring element type and a point on the line represents the average neighbor concentration for a given atomic volume. On the top axis of each subplot, a histogram indicates the relative frequency of the central element at a particular volume. For Type 1 sites, the larger atoms (Nb and Ta) are featured because their segregation to highly compressed sites is unexpected. For Type 2, only Nb and Mo plots are shown since these make up the vast majority of segregating atoms and other elements do not have statistically significant contributions. A complete compilation of local compositions by central element for all coordination types are included in Supplementary Note 3.

Figures 9(a) and (b) illustrate how the decrease in grain boundary structural unit spacing (i.e., the increase in atomic volume due to stress relaxation) influences the density of and local composition around a central atom in compressed sites. For Nb central atoms (Figure 9(a)), high concentrations of Mo neighbors occur at lower volume sites while high concentrations of Nb neighbors occur at higher volume sites. We further observe that the neighboring concentration of Nb atoms increases as atomic volume increases. Not only does reduction in local elastic strain make these larger Type 1 sites more accepting for Nb, but the convergence of grain boundary structural units forces Nb clustering at the back of one kite into close proximity with an adjacent unit's kite tip. Favorable Nb-Nb clustering, generally restricted to expanded defect and near-boundary sites, can then influence the composition of these adjacent sites. On the other hand, Ta atoms are primarily observed in the lower volume Type 1 sites (Figure 9(b)), where Mo neighbors were previously shown to be more abundant. Despite being a highly compressed environment, Ta segregation becomes favorable due to key Mo-Ta ordering, while at volumes greater than 16.5 Å$^3$, the increase in Nb neighbor concentrations makes Ta segregation increasingly inhospitable due, in part, to strong resistance to Nb-Ta interactions.

Type 2 sites for Nb and Mo span a larger range of atomic volumes and are shown in Figures 9(c) and (d). This analysis confirms that the neighbor averages from Figure 8(b) are representative of the majority of local environments, even with large variations in local site volume. For Mo, average neighbor concentrations of Nb and Mo are almost exclusively above and below 0.6 and 0.2 across the entire spectrum of Type 2 sites. Nb central atoms have more balanced concentrations of Nb and Mo neighbors, with a tendency to form more Nb-Nb bonds as site volume increases. Exploration into enthalpy driven interactions in NbMoTaW variants can provide further insight into this behavior. For NbMoTaWV, Fernandez-Cabellero et al.[45] modeled CSRO characteristics to show strong pair-wise interactions between Ta and Mo at low temperatures, but also found that removing Ta led the most favorable interaction at 0 K to become Mo and Nb ordering. Our CSRO analysis from Figure 1(a) and (b) also indicates a slight tendency for Nb-Mo CSRO and strong Mo-Mo repulsion, although both are likely influenced by the strong signal from Mo-Ta interactions in this alloy. The segregation of Mo to Type 2 sites is apparently stabilized by large amounts of vicinal Nb neighbors and is therefore a chemical effect.

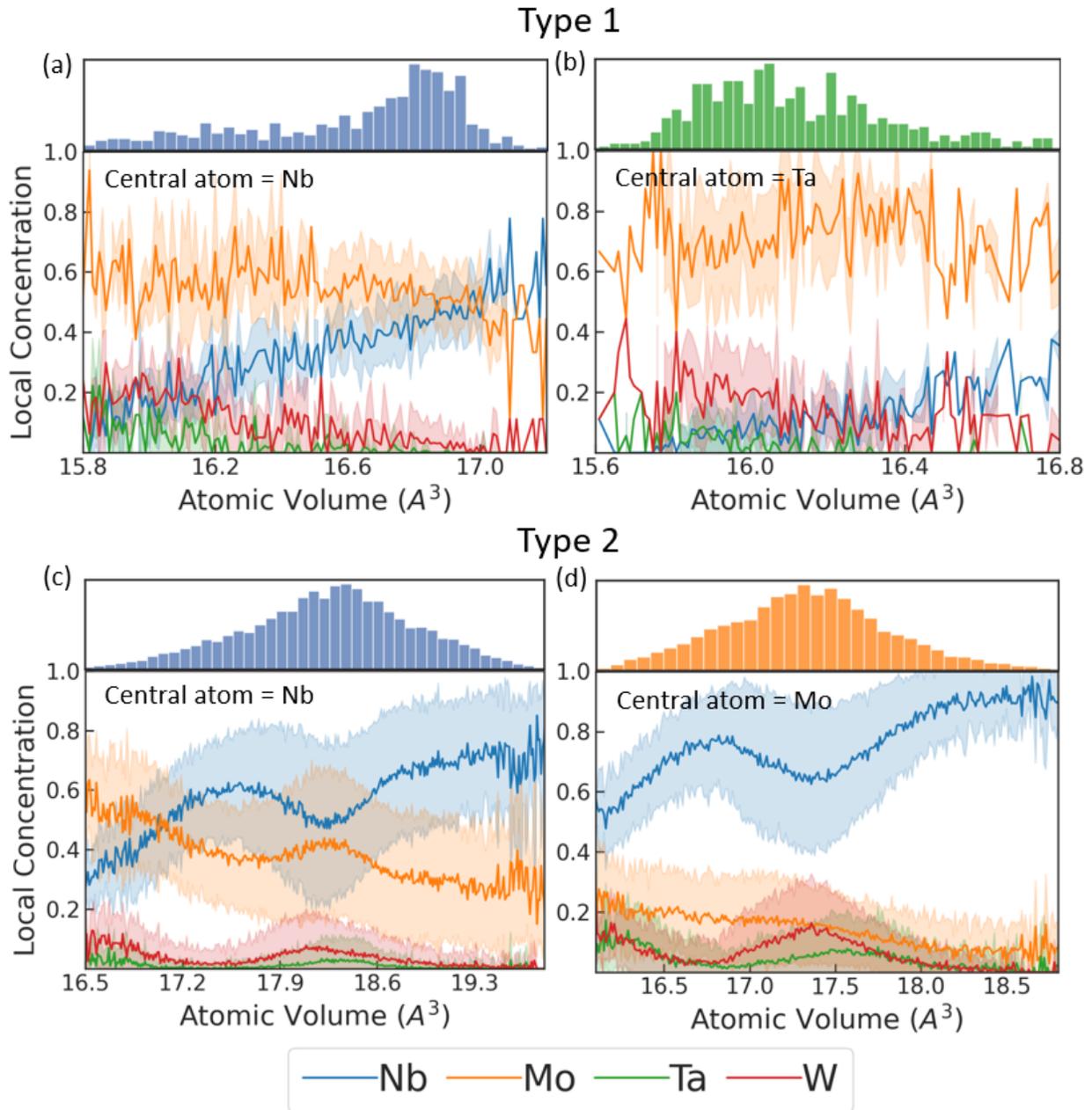

FIG 9. Local concentrations of nearest neighbor elements around a central atom, presented as a function of the central atom's atomic volume. For Type 1 sites, (a) Nb and (b) Ta centered atoms are shown. The top

panel shows the histogram of central atom counts for a given atomic volume, and the bottom panel indicates nearest neighbor concentration profiles as a function of atomic volume. The central atoms Nb and Ta are analyzed because they are the larger atoms in the system unexpectedly observed in highly compressed sites. For Type 2 sites, the same format of data is presented in (c) and (d). The Nb and Mo are shown as these are the primary segregating species (only small amounts of Ta and W are observed in these sites).

To better understand how these critical driving forces for segregation work together, we discuss our results in the context of a binary theoretical model that aims to predict the segregation enthalpy based on the change in local atomic conditions from an initial to final state. Previous work from Wynblatt and Ku,[59] later adapted by Wynblatt and Chatain,[37] organizes the components of segregation enthalpy into a single expression for surface and interfacial segregation in binary alloys as:

$$\Delta H_{seg} = (\gamma_B - \gamma_A)\Omega + 2\omega_{AB}\left[z^l(x^b - x^i) + z^v(x^b - 0.5)\right] - F(M)\Delta r^2 \qquad (4)$$

In Equation (4), segregation enthalpy is composed of three energetic driving forces. The first term is the interfacial energy term, where $\gamma_i$ is the interfacial energy for the $i^{th}$ component and $\Omega$ is the molar area. This term is reflective of the relative difference in cumulative bond energies between the two elements for a given site and is negative when the interfacial energy is lowered by atom B replacing atom A. The second term is the alloy interaction term, where $\omega_{AB}$ is a regular solution constant and is proportional to the enthalpy of mixing of the binary AB alloy, $x^b$ and $x^i$ are the respective bulk and interfacial compositions, and $z^l$ and $z^v$ are the number of bonds of atoms that lie within the same plane or in adjacent planes, respectively. The final term is the elastic solute strain energy term, where $F(M)$ represents a function of the elastic moduli and $\Delta r$ is the difference in atomic radii between the segregating species and the average atomic radii of the bulk. For favorable segregation, $\Delta H_{seg}$ will be negative. Table II provides per-element values that can be used to determine when the different interaction terms should have significant influence. In the first three columns of Table II, interfacial energies for each element at several interfaces indicate Nb as the preferred segregant, with the relative interfacial energies of elements at all selected boundaries following the relationship: Nb < Ta < Mo < W.[60] These values are representative of an element's tolerance to exist in disturbed local bonding environments and the elemental trend derives from the positive correlation between a pure element's bond energy and melting temperature.[61] In this case, the melting points, and therefore the bond energies, have the same relationship as the interfacial energies.[40] Hence, as coordination number decreases, relative differences in bonding energies between elements are expected to play an increasingly influential role in the segregation enthalpy, leading to dominant Nb segregation in heavily distorted, Type 3 sites.. In the final two columns of Table II, the atomic radii computed from a single crystal MPEA model are presented. None of the individual atomic radii differ from the MPEA average (1.594 Å) by more than 0.005 Å. However, the negative values for Mo and W indicate some preference towards compressed sites in order to reduce the local elastic strain, while positive values for Nb and Ta suggest preference for expanded sites.

Based on the CSRO values and the nearest-neighbor relationships in Figures 8 and 9, a critical discussion about the predictive power of Equation (4) can be made on a per-element, per-coordination region basis. For example, we previously observed that roughly 70% of Ta atoms are found in the over-coordinated sites of Type 1; in this case, neither the interfacial energy term nor the elastic strain energy term contributes to favorable segregation enthalpy, suggesting that the high concentration of Mo in the nearest neighbor shell is the primary driver pulling Ta atoms into these sites. We further observed considerable Mo segregation to the Type 2 grain boundary sites despite relatively high interfacial energy and unfavorable elastic strain energy conditions. Here, the high concentration of Nb atoms in the first nearest neighbor shell (Figure 9(d)) indicates that Nb-Mo pairs at the interface are the result of co-segregation. Nb enrichment for all site types is not surprising, having favorable conditions for each term in Equation (4), with additional chemical driving force for segregation coming from Nb-Mo ordering and Nb-Nb clustering. As such, the tendency for Nb to outcompete Ta for similar grain boundary sites can be resolved by considering both the propensity for Nb-Nb bonds and the relative difference between their interfacial energies. Using this framework and the analysis in previous sections, one can observe how local chemistry drives segregation despite structural opposition in some cases, highlighting the importance of chemistry in MPEAs.

|  | Grain Boundary Energies (J/m$^2$) | | | Atomic Radii (Å) | |
| --- | --- | --- | --- | --- | --- |
| **Elements** | Σ3 (011) | Σ5 (001) | Σ7 (111) | Computed in MPEA single crystal | Difference from MPEA average (1.594) |
| Nb | 0.294 | 1.492 | 1.343 | 1.599 | 0.005 |
| Mo | 0.503 | 2.432 | 2.21 | 1.590 | -0.004 |
| Ta | 0.334 | 1.925 | 1.558 | 1.596 | 0.002 |
| W | 0.714 | 3.205 | 2.866 | 1.590 | -0.004 |

**TABLE II.** The first three columns show the elemental grain boundary energies for selected boundaries calculated via DFT.[60] The last two columns are calculated values for atomic radii for the respective elements in the MPEA single crystal, followed by the difference from the bulk average lattice parameter.

### C. Relative structural and chemical contributions to segregation

We next employ a machine learning model to demonstrate and quantify the relative importance of various features for predicting segregation. Similar methodologies have been implemented successfully to predict segregation energies in binary alloys. For example, Messina et. al.[62] used just three structural descriptors (Voronoi volume, coordination number, and hydrostatic pressure) to predict the segregation energies of Al to 14,000+ sites at or near Mg boundaries in a dilute alloy. While the most successful model implemented by these authors integrated a neural network with a decision-tree based algorithm to achieve a root-mean-square error (RMSE) of 8.56 meV, a less sophisticated random forest (RF) algorithm[63] was found to have similar accuracy. Huber et. al.[64] used gradient-boosted decision trees to model various elemental segregation energies in an Al solvent using a larger set of structural descriptors containing bond-orientational order parameters.[65] Greater RMSE values were reported by these authors (40-130 meV), likely due to additional factors including the effects of finite temperature. Importantly, analysis of the feature importance revealed meaningful physical justification for model performance in different solutes. Here, we perform analysis of different local descriptors on segregation behavior using a similar approach. Using a RF algorithm, we create two models for each element based on two different feature vectors as input and atomic potential energy as output. One key difference in our approach from the others is that we use local descriptors from only grain boundary atoms in the fully relaxed interfacial state rather than using unrelaxed grain boundary and bulk site descriptions. While this may lessen the absolute predictive power of the models, our goal is instead to extract key relative differences in model performance and local descriptor impact to understand structural and chemical contributions to segregation.

Two feature vectors were first constructed composed of different components of the local atomic environment to understand how their respective environmental descriptors could be used to predict the potential energy of a grain boundary atom. In the first feature vector (Model 1), only structural descriptors were considered including the effective coordination number, Voronoi volume, and hydrostatic pressure. In addition to these local structural features, the second feature vector (Model 2) included chemical information in the form of nearest-neighbor local elemental concentrations with respect to the central atom. For each element, two models were developed using these input vectors to both understand the relative predictive power of each feature set and to distinguish the individual contributions of structural and chemical effects on local site energies within the boundaries. The RF algorithm type was selected due to its ease-of-use, computational efficiency,[66] and its successful application to grain boundary segregation in past studies.[62,64] The respective number of total atoms for each subset are listed in the last column of Table S1 and correspond only to defect atoms. To optimize and train the models, a train/test/validation split of 70%/15%/15% was utilized. Hyperparameter tuning was performed using five-fold cross-validation and a randomized search method on the validation set. Subsequent training was performed on the train group, and the final model was evaluated on the test group. The following discussion focuses on Nb and Mo segregants, as these elements provide enough data points for a robust analysis, however the model results for Ta and W are also included in Figure 10 and similar results are achieved.

The results for each model and element are shown in Figure 10. RMSE and coefficient of determination ($R^2$) values are provided for each regression and a solid line indicates a perfect fit. Two dashed lines indicate a ±0.5% deviation (total width of 1%) of the predicted value from the true potential energy. Model 1 (Figures 10(a) and (b)) produces better $R^2$ and RMSE values for Nb (0.617 and 80 meV) as compared to Mo (0.512 and 96 meV). Model 1 has significantly lower accuracy compared to the results from Messina et al.,[62] but this is not wholly unexpected as we have shown that chemical interactions become very important in the MPEA. Furthermore, we have shown above that BCC/bulk sites can influence segregation, but such a model cannot account for this effect fully. When including the first shell concentration descriptors in Model 2 (Figures 10(e) and (f)), a very large improvement in model accuracy is achieved for Mo, with a 37.7% increase in $R^2$ and a 22.5% decrease in RSME. Smaller but still significant improvements are achieved for the Nb model, with a 17.7% increase in $R^2$ and a 11.3% decrease in RMSE. We note that the improvement of these metrics with added chemical information is of primary importance, rather than their absolute value.

Gradient coloring is used within each plot to show the point density for each model. When adding chemical descriptors to the RF model, there is significantly improved alignment of the data. Most notable is the much-improved alignment of high energy points along the top right area of the perfect fit line for Mo when comparing Model 2 to Model 1. Higher energy sites for Mo generally correspond to sites with lower coordination and predominantly have Nb neighbors (Figure 6(a) and Figure 9(d)), thus showing the impact of including local concentration information in Model 2. A similar but less dramatic improvement is observed for Nb at the lower energy sites on the bottom left. One can also use the (admittedly arbitrarily chosen) error bounds of ±0.5% to quantify this improved alignment. For the improved Model 2, Mo has approximately 11% more points within this interval while Nb has approximately 6% more points included within the interval.

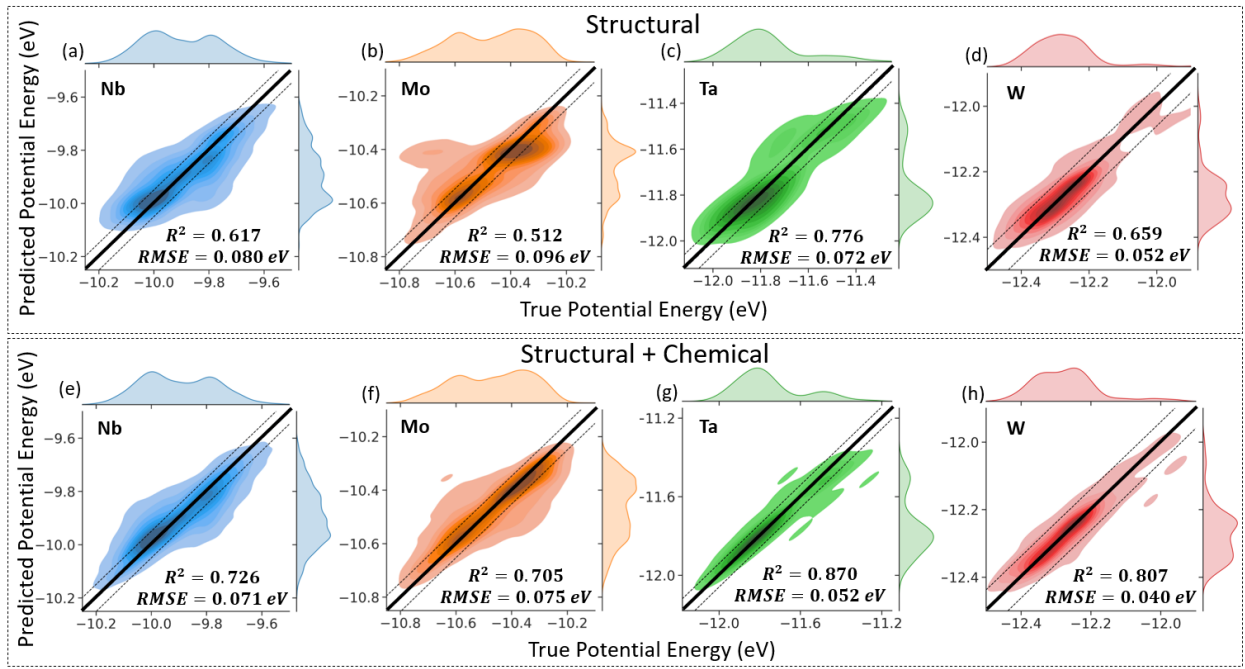

**FIG 10. Results of RF prediction are shown using two sets of feature vectors. Headers in each dashed box indicate the set of feature vectors used to model the potential energy predictions. The solid line in each plot corresponds to a perfect fit, while the dashed lines represent ±0.5% deviation from perfect fit to either side (1% total width). Included in each plot are the R-squared and RMSE values. Joint axes and gradient coloring indicate point density for each plot.**

To better understand the individual contribution of each descriptor in Model 2, we calculate and compare their relative feature importance in Figure 11 through a metric known as the Gini importance.[67] This value provides relative comparisons between input parameters and their fractional effect on model predictions. For both Nb and Mo, the three structural descriptors dominate, accounting for approximately 82% and 76% of the total feature importance for Nb and Mo, respectively. However, the remaining 18% and 24% contributions from chemical factors are indeed significant and must be accounted for. Analysis of individual neighboring elements reveals that Nb is most strongly affected by Nb neighbors (~12% of total effect) while the Mo model is influenced roughly equally by Nb and Ta neighbors (~7% of total effect for each). In both cases, the dominant chemical feature aligns with expected bulk CSRO behavior observed in Figures 1(a) and (b), indicating that these interactions play an important role not only in bulk behavior, but in segregation behavior as well. Similar magnitudes of importance are observed for chemical descriptors in the Ta and W models. Finally, we reiterate that these machine learning analysis models are not fully optimized and are not meant to be seen as the ideal set of descriptors or modeling frameworks. This analysis aims to isolate the individual contributions of structure and chemistry as much as possible, building on well-established models.

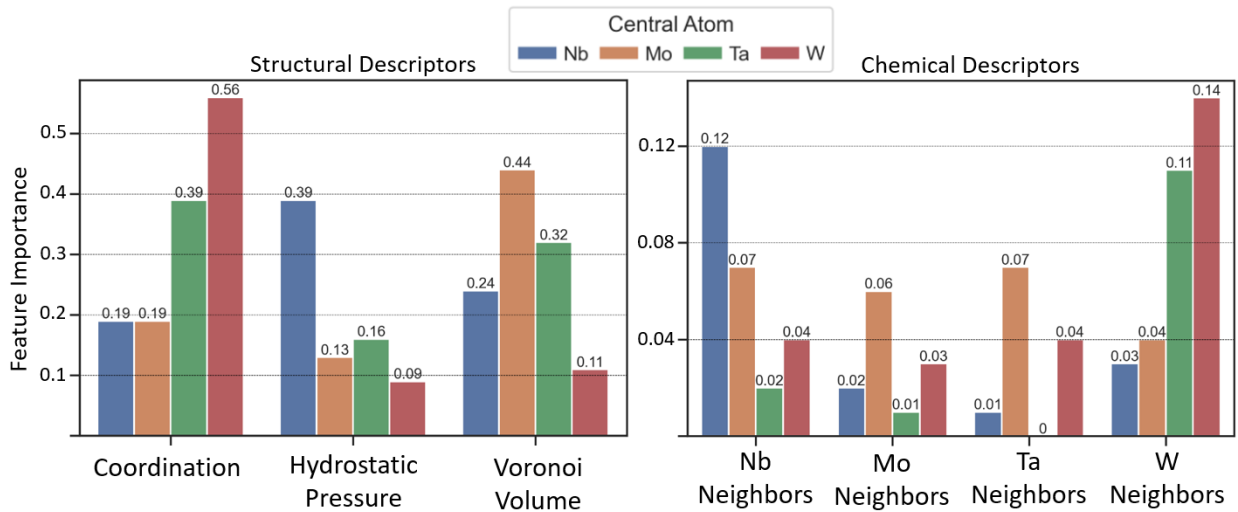

**FIG 11. The first plot titled Structural Descriptors shows the relative importance of each of the three local structural features used in Model 2, including the coordination number, the hydrostatic pressure, and the Voronoi volume. The second plot titled Chemical descriptors provides the relative importance of the corresponding local chemical features used in Model 2, consisting of the concentrations of each element in the nearest neighbor shell. For each feature, the colors of the bars represent the central element used in the model.**

## IV. CONCLUSIONS

Using atomistic simulations on hundreds of boundaries and over a hundred thousand grain boundary sites across the [001] symmetric tilt axis in equiatomic NbMoTaW, interfacial segregation behavior was observed to be driven by a complex interplay between local structure and chemistry in MPEAs. Our work here allows for the following conclusions to be made:

- Strong segregation behavior was observed at all angles with some compositional variations in different angular regions. Disconnected structure regions at low and high tilt angles exhibit the most compositional variation for both defect and adjacent BCC atoms. In these regions, Ta and W deplete incompletely, indicating special structural or chemical influence at specific sites. In the continuous defect structure boundaries, Nb enrichment is the greatest with Mo filling most of the remaining sites.

- Using effective coordination to isolate similar grain boundary site types, trends in elemental tendencies towards specific sites are shown, with Type 1 sites having the most compositional diversity. Particularly, Ta segregation to over-coordinated/compressed sites suggests elemental pinning by nearby Mo neighbors and is confirmed by local chemical analysis. For Type 2 sites, Mo and Nb are predominantly observed despite having distinctly different local compositional profiles. We show that Mo favors Nb interactions when Ta is not present, encouraging co-segregation behavior with Nb to enable segregation to structurally unfavorable sites. Further Mo segregation is likely limited due to unfavorable Mo-Mo interactions. In contrast, Nb has an affinity for itself and for Mo, favoring clustering and ordering behavior simultaneously and promoting stronger segregation.

- Using a machine learning framework, several models were implemented to isolate the contributions of structural and chemical driving forces on segregation behavior. Without any chemical information, Nb is predicted more successfully than Mo, suggesting that Nb segregation has a larger structural driving force of the two. With both structural and chemical descriptors used as input parameters, the prediction of Mo segregation shows the most improvement, demonstrating a larger influence of local chemistry on its segregation behavior. Subsequent analysis of feature importance confirms the impact of chemical interactions in the context of CSRO tendencies.

As a whole, the analysis presented here fills a gap left by prior atomistic segregation studies of NbMoTaW polycrystalline samples.[27] That is, while segregation behavior in this alloy follows a predictable structural trend overall, unexpected solute adsorption to specific sites can be attributed to favorable local chemical environments. These findings suggest that pathways towards microstructural engineering exist in MPEAs and can be tailored through their elemental complexity. With interfacial structure largely determining the grain boundary composition, future experimental studies should look to confirm differences in segregation behavior for different crystallographically oriented boundaries at temperatures (i.e., < 1000 K) where CSRO is predicted to be a strong effect. Furthermore, our results suggest that changes to interfacial composition can be achieved by altering global element concentrations and therefore bulk ordering behavior, which may have substantial influence on grain boundary properties such as mobility, nanocrystalline stability, and phase transitions. The identification of these unique segregation events requires the study of a large variety of grain boundary sites, which was accomplished in this work with high-throughput modeling using both low energy and metastable grain boundaries as templates for segregation.

**SUPPLEMENTARY MATERIAL**

See supplementary material for additional details on atomic counts by coordination regime, elemental distributions by atomic volume for Type 1 sites and analysis of local concentrations as a function of atomic volume for central atoms in their respective coordination regimes.

## ACKNOWLEDGEMENTS

This research was primarily supported by the National Science Foundation Materials Research Science and Engineering Center program through the UC Irvine Center for Complex and Active Materials (DMR-2011967). The authors thank Prof. Shyue Ping Ong for the helpful discussions regarding machine learning interatomic potential utilized in this work.

## AUTHOR DECLARATIONS

**Conflict of Interest**

The authors have no conflict of interest to disclose.

## DATA AVAILABILITY

The data that support the findings of this study are available within the article and its supplementary material.